\documentclass[%
aip,
amsmath,amssymb,
reprint,%
]{revtex4-1}
\usepackage{braket}
\usepackage{graphicx}
\usepackage{dcolumn}
\usepackage{bm}

\usepackage[utf8]{inputenc}
\usepackage[T1]{fontenc}
\usepackage{mathptmx}
\usepackage{placeins}  
\usepackage{bibentry}

\usepackage{amsmath}
\usepackage{amsfonts, bbm, accents}

\usepackage[usenames,dvipsnames]{color}
\usepackage[breaklinks=true]{hyperref}
\hypersetup{
  colorlinks   = true, 
  urlcolor     = blue, 
  linkcolor    = blue, 
  citecolor   =  MidnightBlue 
}

\usepackage{soul} 

\usepackage[capitalise]{cleveref} 
\crefformat{equation}{Eq.~(#2#1#3)} 
\crefformat{section}{Sec.~#2#1#3} 
\Crefformat{equation}{Equation~(#2#1#3)}
\crefformat{figure}{Fig.~#2#1#3}
\crefrangeformat{equation}{Eqs.~#3(#1)#4--#5(#2)#6}
\Crefformat{section}{Section~#2#1#3}

\begin{document}


\title{Perspective: Reproducible Coherence Characterization of Superconducting Quantum Devices}

\author{Corey Rae H. McRae}
\email[]{coreyrae.mcrae@colorado.edu}
\affiliation{
Department of Physics, University of Colorado, Boulder, Colorado 80309, USA}
\affiliation{ 
National Institute of Standards and Technology, Boulder, Colorado 80305, USA
}%
\affiliation{ 
Boulder Cryogenic Quantum Testbed, University of Colorado, Boulder, Colorado 80309, USA
}%
\author{Gregory M. Stiehl}
\affiliation{
Rigetti Computing, Berkeley, California 94710, USA}
\author{Haozhi Wang}
\affiliation{
Laboratory for Physical Sciences, University of Maryland College Park, College Park, MD 20740, USA
}%
\author{Sheng-Xiang Lin}
\affiliation{
Department of Physics, University of Colorado, Boulder, Colorado 80309, USA}
\affiliation{ 
National Institute of Standards and Technology, Boulder, Colorado 80305, USA
}%
\affiliation{ 
Boulder Cryogenic Quantum Testbed, University of Colorado, Boulder, Colorado 80309, USA
}%
\author{Shane A. Caldwell}
\affiliation{
Rigetti Computing, Berkeley, California 94710, USA}
\author{David P. Pappas}
\affiliation{ 
National Institute of Standards and Technology, Boulder, Colorado 80305, USA
}%
\author{Josh Mutus}
\affiliation{
Rigetti Computing, Berkeley, California 94710, USA}
\author{Joshua Combes}
\affiliation{
Department of Electrical, Computer, and Energy Engineering, University of Colorado, Boulder, Colorado 80309, USA}

\date{\today}

\begin{abstract}
As the field of superconducting quantum computing approaches maturity, optimization of single-device performance is proving to be a promising avenue towards large-scale quantum computers. However, this optimization is possible only if performance metrics can be accurately compared among measurements, devices, and laboratories. Currently such comparisons are inaccurate or impossible due to understudied errors from a plethora of sources. In this Perspective, we outline the current state of error analysis for qubits and resonators in superconducting quantum circuits, and discuss what future investigations are required before superconducting quantum device optimization can be realized.
\end{abstract}


\maketitle 

Superconducting quantum computing is poised to become a successful platform for large-scale quantum computing due to promising qubit performance, ease of multiqubit coupling, and potential for scalability.~\citep{Gambetta2017} Currently, superconducting quantum computing is transitioning to a mature field of research with efforts from large industrial and governmental organizations as well as many academic groups. Architectures of more than 50 qubits have already been demonstrated.~\citep{Arute:2019aa} 
However, cryogenic single-photon microwave losses, which limit coherence of individual qubits and resonators, are a performance bottleneck for these systems,~\citep{Muller2019, McRae2020} and efforts to mitigate these losses have been curtailed by understudied errors and unreported experimental details that restrict the accuracy of interlaboratory comparisons.

\begin{figure*}
\includegraphics[width=160mm]{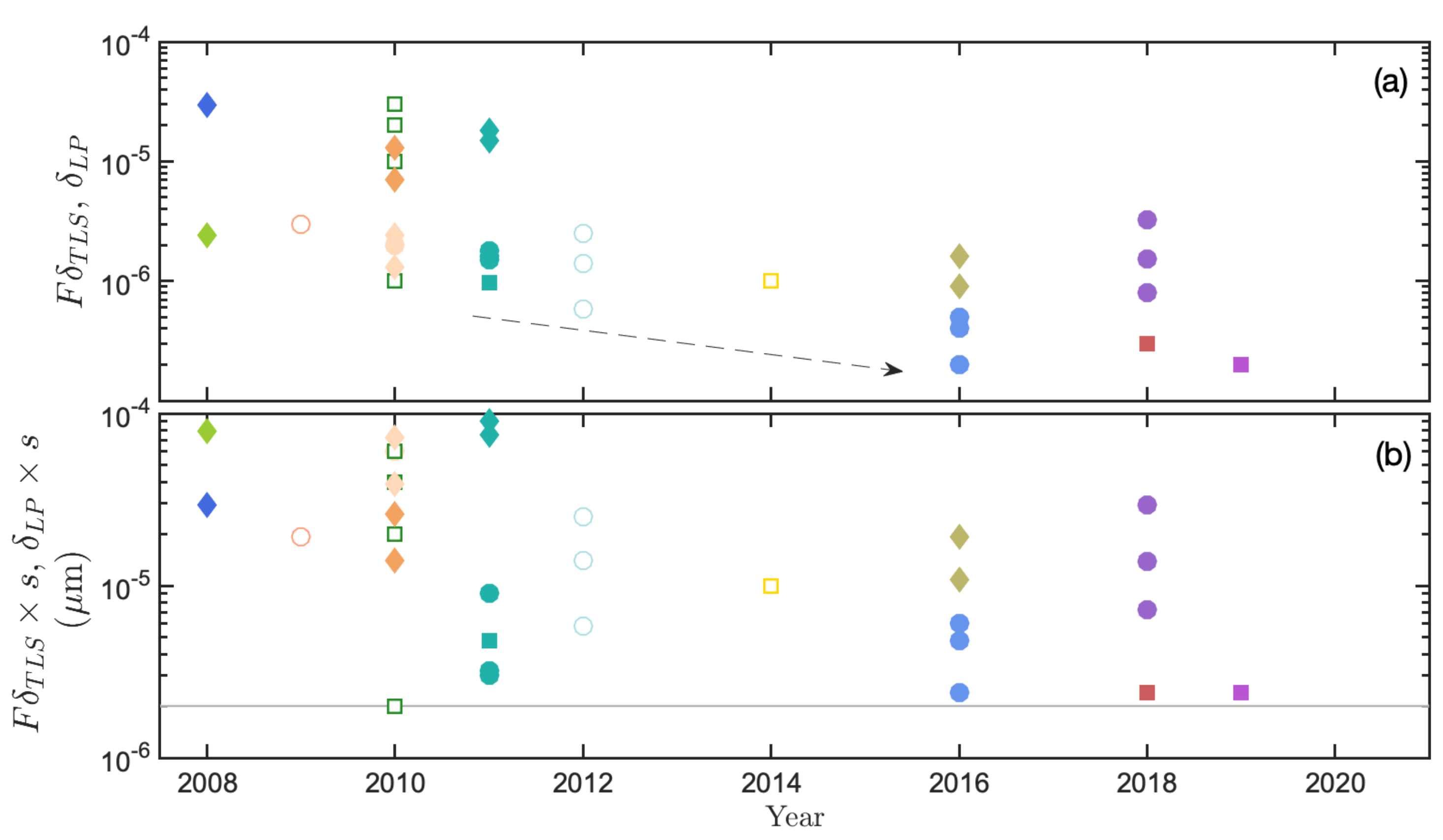}
\caption{\label{fig:motivation}Summary of thin film superconductor loss measurements as a function of publication year. Note: Values are not directly comparable due to differences in resonance frequency, $Q_i/Q_c$ matching, fabrication, shielding, microwave chain, thermalization, and data fitting. (a) Coplanar waveguide (CPW) resonator loss of the best performing "hero" devices, and (b) the same loss multiplied by CPW gap width for normalization purposes. Solid markers: resonator-induced intrinsic TLS loss $F \delta_{\mathrm{TLS}}^0$ and $F \delta_{\mathrm{TLS}}^0 \times s$, where $s$ is CPW resonator gap width in $\mathrm{\mu}m$. Open markers: $\delta_{\mathrm{LP}}$ and $\delta_{\mathrm{LP}} \times s$ values, where $\delta_{\mathrm{LP}}$ is low power loss. Superconducting microwave resonators patterned from Al (circles), TiN (squares), and Nb (diamonds) are shown. Marker color denotes reference in which each measurement is reported, with associated colors and references shown in Ref.~\citenum{McRae2020}. Original data can be found in Refs.~\citenum{Wang2009,Macha2010,Sage2011,Megrant2012,Richardson2016,Earnest2018,Vissers2010,Sage2011,Ohya2014,Calusine2018,Lock2019,Gao2008b,Kumar2008,Wisbey2010,Goetz2016}. Grey arrow is a guide to the eye showing a decrease in measured loss over time. Grey line denotes the highest performance seen for on-chip superconducting microwave resonator materials loss.}
\end{figure*}

\begin{figure}
\includegraphics[width=80mm]{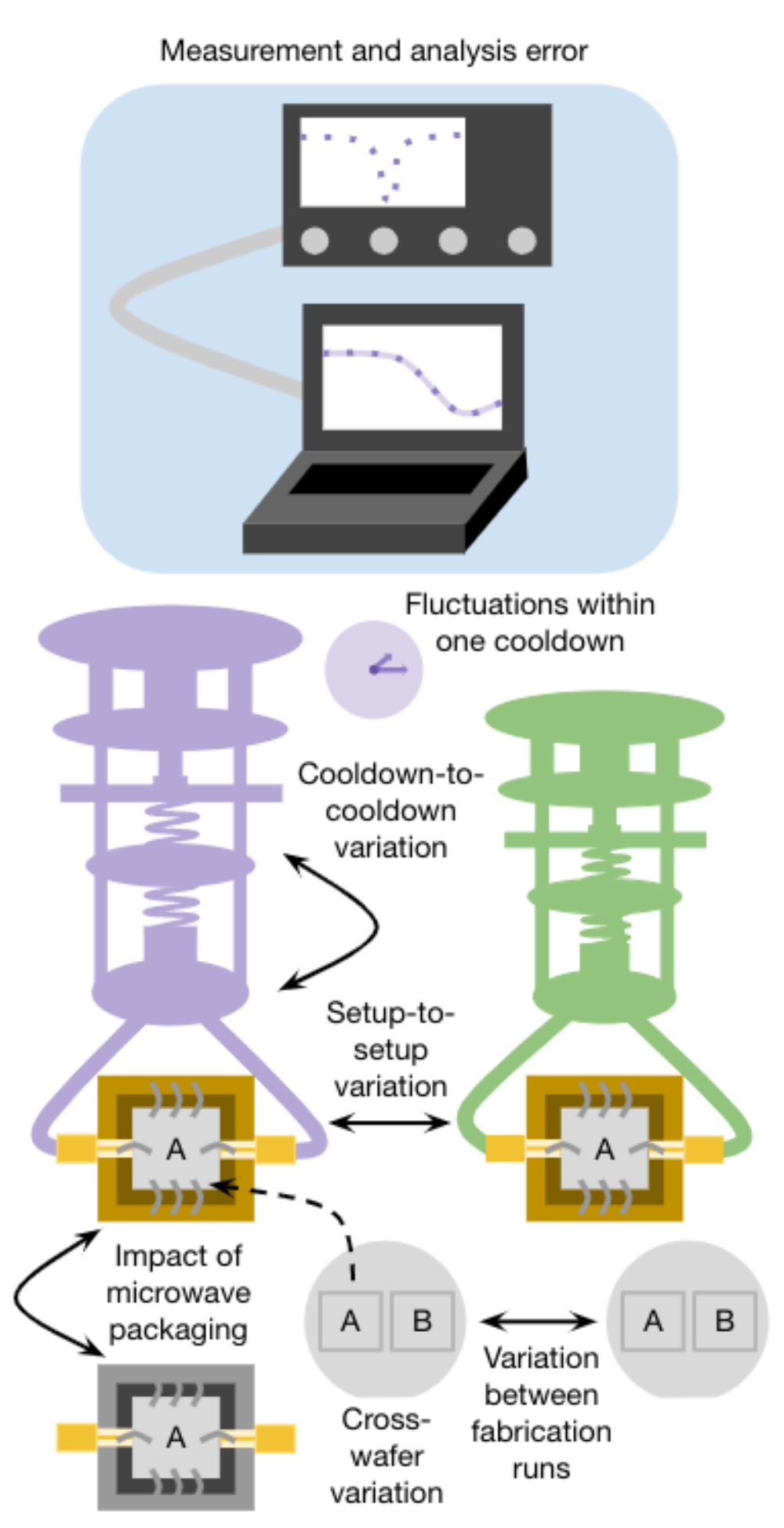}
\caption{\label{fig:fluctuationoverview}Schematic of the seven categories of error at play in a superconducting quantum circuit measurement. Two dilution refrigerator (DR) setups are shown in purple and green, two sample boxes with wirebonds are shown as gold and grey boxes, and two wafers from different fabrication runs are shown as light grey and dark grey circles with small cord cuts.}
\end{figure}

The optimization of individual device performance requires the ability to accurately and precisely compare performance metrics both within a single lab (i.e., A/B device comparisons) and among labs (i.e., comparisons of published measurement outcomes). However, previous experiments have shown that device performance fluctuates in time and among experimental setups as a function of a variety of parameters, many of which are poorly understood. As well, efforts to increase device performance have historically drawn conclusions from the experimental characterization of a small number of devices, and sometimes a single measurement of a single device, due to the large investment of time and money required for a single device measurement.

At present, it is hard to argue that a particular experiment has truly seen a difference in performance, or whether the results are effectively unchanged within the uncertainty of the experiment, since the magnitude of uncertainty in superconducting quantum device performance measurements is largely unknown.

Consider a simple question: Does a certain surface treatment increase or decrease device performance? In theory, two identically-designed circuits would be fabricated, one of which is exposed to the surface treatment and one which is not. Then, the samples would be cooled down in a dilution refrigerator to $\sim$ 10 mK and performance would be measured. The responses would be analyzed, and the performance metrics would be reported for each device. If the performance of the treated sample is better than the untreated sample, we can conclude that the sample treatment increased the device performance. If the difference in values is within the uncertainty of those values, then we can conclude that the effect of the surface treatment is smaller than the precision of the measurement or is negligible.

In reality, these comparisons are not so simple. While the fitting uncertainty of performance metrics for qubits and resonators is generally knowable and small, the metrics themselves vary significantly within a single cooldown,~\citep{Burnett2019,Schlor2019,Klimov2018} from cooldown to cooldown,~\citep{Burnett2019,Schlor2019,Calusine2018} among nominally identical devices on a single chip or wafer,~\citep{Burnett2019,McRae2021,Calusine2018} and among nominally identical devices from different fabrication runs.~\citep{Woods2019} The energy relaxation time $T_1$ of a single qubit in a nominally fixed environment can vary by $\sim$50$\%$ over a period of hours.~\citep{Klimov2018} So before we can address the question of which surface treatment is better, we must first address the question: What is the uncertainty of our reported performance metrics, and how can we minimize this uncertainty to facilitate performance metric comparisons while still allowing the experiment to be practically implementable?

Fig.~\ref{fig:motivation} shows the importance of these types of A/B device comparisons to the future of superconducting quantum computing as a research field. In Fig.~\ref{fig:motivation}~(a), resonator loss is plotted as a function of year of publication for the three most common superconductors: Al, TiN, and Nb. A downwards trend in loss is seen for the highest-performing devices, indicating that the field is progressing toward devices with lower loss and thus higher performance. However, when device design is accounted for by multiplying by the coplanar waveguide resonator gap $s$ in order to estimate the filling factor of the TLS material (Fig.~\ref{fig:motivation}~(b)), this trend disappears and we see no significant trend in progress in materials loss over the past decade, indicating that the increasing performance trend in (a) could be attributed to design changes. This demonstrates that materials loss mitigation may truly be an area of untapped potential in the effort towards large-scale superconducting quantum computing. On the other hand, the scatter and lack of error bars on these data points indicate a fundamental flaw in experimental design in the field: the accuracy and precision of loss measurements have historically been insufficient to drive improvement.

In this Perspective, we will outline the standard performance metrics and what is currently known about fluctuations in performance for superconducting qubits and resonators. We will also highlight areas of uncertainty that must be filled in before optimization of superconducting quantum device performance can be achieved. This article is not intended as a review or tutorial for superconducting qubits or resonators; readers interested these types of articles are directed towards Refs.~\citenum{Krantz2019, gao_practical_2021,McRae2020,Zmuidzinas2012}. 

The major sources of error and performance fluctuation that hamper reproducibility studies are schematically depicted in Fig.~\ref{fig:fluctuationoverview}. Conceptually, we can organize these into groups of sources that apply to different experiments. For example, the comparison of performance of a device before and after some materials treatment requires accounting for measurement and analysis error (Sec.~\ref{sec:measerror}), fluctuations within one cooldown (Sec.~\ref{sec:time}), and cooldown-to-cooldown variation (Sec.~\ref{sec:cooldown}). An interlaboratory comparison of a single device requires that all above error sources as well as setup-to-setup variation (Sec.~\ref{sec:setup}) be addressed. For a comparison of devices on two separate chips, the above error sources as well as the impact of microwave packaging (Sec.~\ref{sec:mwpackage}) and variation due to fabrication (Sec.~\ref{sec:faberror}) will play a part.

In this work, we assume 2D transmon and 2D resonator implementations, the two primary components of planar superconducting quantum circuits. However, much of the error analysis described here applies to other superconducting quantum devices as well. It is the hope that this Perspective will facilitate measurements of metrics that are both self-replicable and reproducible among laboratories. 

\section{Introduction to Performance Metrics}\label{sec:metrics}
This section outlines the common performance metrics used to characterize qubits and resonators. 
Performance metrics of qubits and resonators are affected not only by intrinsic sources such as materials losses but also by aspects of the experimental setup such as microwave components and thermalization, as well as the data analysis process.
Here, we assume measurements occur in the quantum computing regime of $\sim$ 10 mK temperatures, single photon powers, and zero applied magnetic field, unless otherwise stated.

\subsection{Resonator Performance Metrics}

The internal quality factor $Q_i$ of a resonator is the ratio of resonator energy stored over energy lost to the environment per cycle and can be determined from a measurement of the scattering parameters. The internal quality factor is inversely related to loss $\delta$ as $Q_i^{-1} \simeq \delta$, and loss from various sources are linearly additive. Due to this additive property as well as the variety of power- and temperature-dependent behaviors exhibited by loss sources, resonator loss is a convenient tool for distinguishing between loss sources. In particular, we can determine the contribution to loss from two-level systems (TLS), the dominant source of loss in state-of-the-art superconducting quantum circuits. 

The intrinsic TLS loss $\delta_{TLS}^0$ is also used as a performance metric.~\citep{McRae2020} A lower value of $\delta_{TLS}^0$ indicates better performance materials and/or interfaces. $\delta_{TLS}^0$ can be extracted from power-dependent resonator measurements provided the geometry-dependent filling factor $F$~\citep{DeGraaf2017} of TLS--contributing material is known. In practice, $F$ is difficult to determine due to design complexity, so the product $F \delta_{TLS}^0$ is reported instead. This is done by sweeping power, i.e., the average photon number in the resonator, and fitting to the total loss.~\citep{GaoThesis,Richardson2016} More information on accurate loss measurements of superconducting microwave resonators can be found in Ref.~\citenum{McRae2020}.

\subsection{Qubit Performance Metrics}

The performance of a single superconducting qubit is typically characterized by the characteristic times for energy relaxation $T_1$ and transverse relaxation $T_2$ or pure dephasing $T_\phi$, where $1/T_{2} = 1/2T_1 + 1/T_\phi$. The longer these times, the better the performance of the qubit. These parameters are measured using well-calibrated single qubit pulses. For a full description of $T_1$ and $T_2$ measurement protocols, see Ref.~\citenum{Krantz2019}. The ability to distinguish between relaxation and pure dephasing is critical to the understanding and eventual mitigation of qubit noise factors. In addition, the value of $T_1$ when the relaxation is dominated by intrinsic materials factors, rather than those induced by the experimental setup, is ideal for reporting in qubit optimization experiments. In this work, we refer to this value as the materials-induced $T_1$.

In order to understand the spectral environment of the qubit in a holistic way, we need to be able to quantify the effects of coherent TLS within some range near the idle qubit frequency. A useful metric would be one which takes into account both the coupling strength $g$ of individual coherent TLS to the qubit as well as the total number of coherent TLS in some probed frequency bandwidth which is ideally the total tunable frequency bandwidth of the qubit. $g$ can be determined for strongly-coupled coherent TLS by measuring the avoided crossing of the transmon with the TLS defect, as shown in Fig.~\ref{fig:SPI}. Here, this information is obtained by performing a qubit spectroscopy experiment with the XY drive as a function of applied magnetic flux to tune the transmon $f_{01}$ transition. When this transition interacts with a nearby defect within the tunability band of the qubit, avoided crossings can be resolved. The avoided crossing splitting is related to the coupling by 2$g$. 

A metric that describes the coupling strength and number of coherent TLS would allow us to place a value on the bandwidth cleanliness of a tunable qubit. By combining time series datasets of $T_1$ and $T_\phi$ in addition to this metric, we can obtain a more descriptive set of metrics for the general coherence of a qubit.

\section{Error Analysis of Performance Metrics}
When comparing two devices or measurements, many sources of random and systematic error contribute to the reported uncertainty. The magnitudes of all of these errors must be taken into account for fair device and process comparisons. Thus, a full accounting of the total error is required in order to understand the factors contributing to variation in performance metrics in superconducting quantum circuits.

As illustrated in Fig.~\ref{fig:fluctuationoverview}, errors can be organized into seven categories: Measurement and analysis error, fluctuations within one cooldown, cooldown-to-cooldown variation, impact of sample box, setup-to-setup variation, cross-wafer variation, and variation between fabrication runs. Depending on the details of the experiment, some or all of these error types may be relevant. For example, an intralaboratory comparison of two devices may not induce setup-to-setup variational errors. In the following sections, a non-comprehensive overview of error of each type is explored for both qubits and resonators. 

\section{Measurement and Analysis Error}\label{sec:measerror}
In order to obtain performance metrics for qubits and resonators, information about the underlying distributions is extracted from a dataset. This process can induce errors in a number of ways.

\subsection{Resonator Measurement and Analysis Error}

Error induced by the vector network analyzer during resonator S-parameter measurements is generally negligible, and fit uncertainty for fitting both $Q_i$ and $F \delta_{TLS}^0$ is knowable and low (see Table~\ref{tab:res} for summary).

Anecdotally, preprocessing and fitting methods have been shown to strongly affect resonator $Q_i$ and $F \delta_{TLS}^0$, with an error magnitude dependent on resonator coupling and impedance mismatch~\citep{Khalil2012} among other factors. Future research is required in this area to pinpoint the low-error measurement regime as well as characterize the error in both optimal and non-optimal measurement regimes.

\subsection{Qubit Measurement and Analysis Error}

Single-qubit pulses are implemented when performing $T_1$ and $T_{\phi}$ characterization. Systematic pulse errors can affect measurement outcomes, but the exponential nature of the decays mean these effects are irrelevant. Moreover, sample-efficient and Heisenberg-uncertainty-limited estimation schemes exist for calibrating such pulses.~\cite{Kimmel_RPE_2015}

Measurements of qubit $T_1$ and $T_{\phi}$ are largely insensitive to changes in analysis as well as minor changes in experimental data acquisition such as modifying the number of points taken to fit to a $T_1$ curve or the number of shots per point, as fitting to an exponential model is more forgiving than the Lorentzian model used to fit $Q_i$. Nevertheless, the estimation of both parameters can be sensitive to drift in the readout quadrature over the time scale of a single experiment.

Errors in potential spectral cleanliness metrics can be induced when identifying TLS and distinguishing them from other signatures in qubit data. The identification of TLS can be performed in multiple ways depending on the data used.

\begin{figure}
\includegraphics[width=85mm]{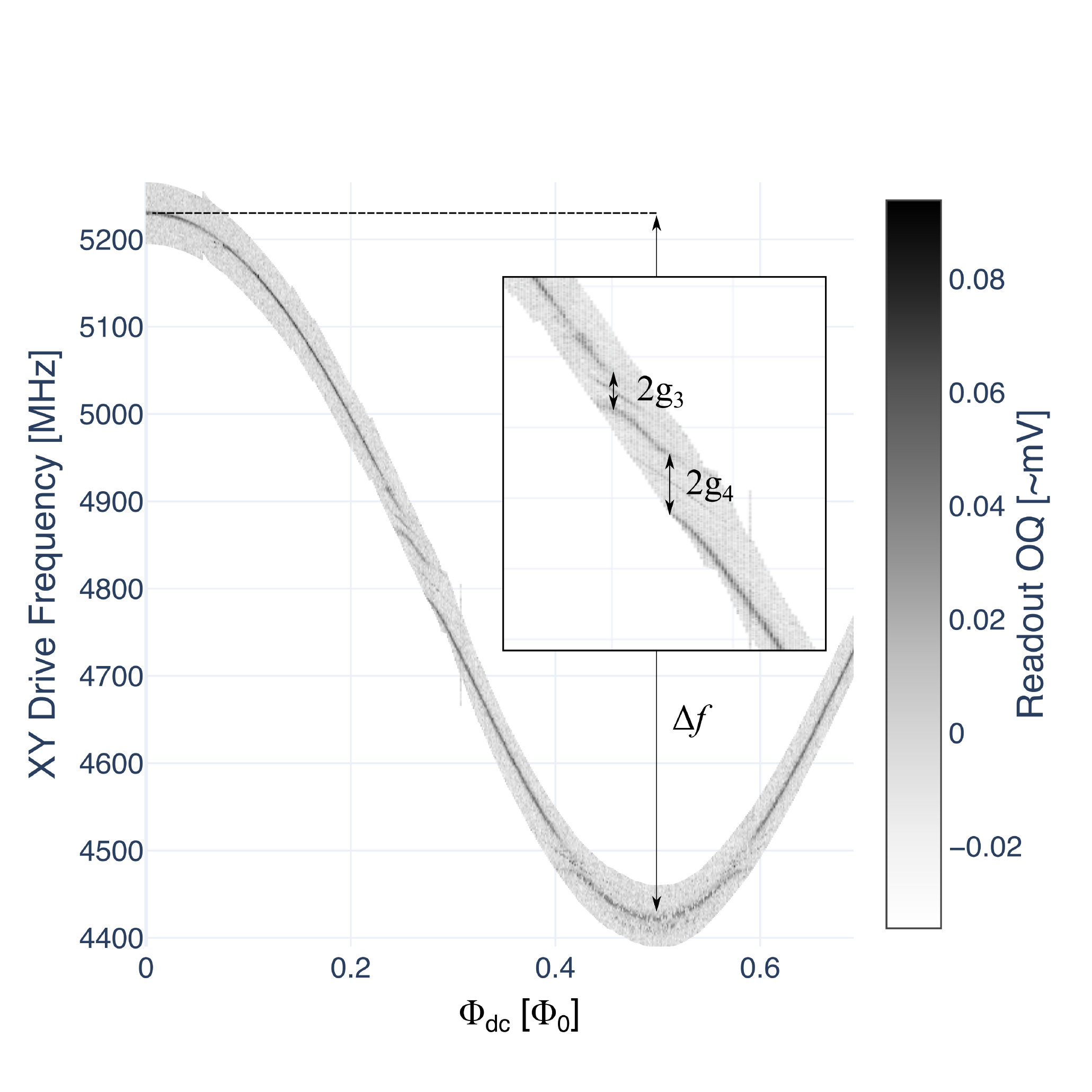}
\caption{\label{fig:SPI}Qubit spectrum as a function of XY drive frequency and applied DC flux for a tunable transmon. Avoided crossings due to the interaction between a qubit and coherent, strongly-coupled TLS are resolved. Five avoided crossings are observed in this spectrum, with $g_1$ = 0.65 MHz, $g_2$ = 0.75 MHz, $g_3$ = 11.4 MHz, $g_4$ = 22.35 MHz, and $g_5$ = 9.8 MHz, over a total tunability band of $\Delta f$ = 800.5 MHz. Inset: zoomed-in plot of two of the avoided crossings ($g_3$ and $g_4$). The population observed in the center of the avoided crossing is the result of a $(f_{\mathrm{qubit}}+f_{\mathrm{TLS}})/2$ multi-photon Bell-Rabi transition.}
\end{figure}

In Ref.~\citenum{Klimov2018}, TLS are identified manually from data such as in Fig.~\ref{fig:KlimovFig} and their $g$ and $f_i$ are determined by fitting to an energy-relaxation model consisting of a sum of Lorentzians, as given in Ref.~\citenum{barends2013coherent}. Alternatively, avoided crossings can be characterized as in Fig.~\ref{fig:SPI}. The automation of these TLS detection processes could help avoid user-induced error.

\section{Fluctuations Within One Cooldown}\label{sec:time}

Within a single cooldown, the performance metrics introduced in \cref{sec:metrics} fluctuate in time with varying time scales. Several mechanisms have been linked to or are suspected to cause temporal fluctuations, including ambient electronics temperature fluctuations~\cite{rigetticomm2,rigetticomm}, TLS loss,~\citep{GaoThesis,Pappas2011} quasiparticle loss,~\citep{Zmuidzinas2012,Serniak2018} and magnetic vortex loss.~\citep{song2009} Regardless of cause, these fluctuations necessitate the use of time series analysis. In this section, we outline some physical mechanisms, as well as discuss the measurement, analysis and characterization of time series data, with the aim to report typical values and understand temporal and spectral correlations.

\subsection{Time Series Analysis}

If some performance metric $P$ changes in time, the value and the uncertainty of $P$ at some particular time is not sufficient to characterize the full behavior of that device. To quantify and analyze fluctuations of a performance metric $P$ within a single cooldown, measurements at many different times $t_n$ are needed, which are samples from a stochastic process $P(t_n)$~\citep{Drascic_2016}. The time between successive samples, known as the sampling period, determines the minimum and maximum frequency component of $P(t)$ that can be estimated from the time series data.

Time series data can then be processed in many ways that illuminate different aspects of the physical sources of fluctuation. Common tools include auto- and cross-correlation functions, power spectral densities,~\citep{kay1988modern} causal models,~\citep{pearl2009causality} process control charts,~\citep{heckert2002handbook,box2005statistics} and Allan variance methods.~\citep{Guerrier2016,Siraya2020} These tools can be diagnostic (e.g. detect and classify an instability as drift or diffusion in the experiment via statistical control charts), qualitative (e.g. power spectral densities~\cite{Klimov2018,Burnett2019,Moeed2019} and Allan variance charts~\cite{Burnett2019,Moeed2019} which can point to certain noise sources). Quantum causal modelling\cite{Costa_2016,QuantumSPcausal} is being applied to gate level characterization of superconducting devices to identify causal structure.~\cite{white_diagnosing_2021}. Using these tools, time series data can be leveraged to understand dynamics and to assess overall performance.


When taking and analyzing time series data, it is important to keep in mind that performance metrics depend on external parameters other than time. For example, the measured $T_1$ of a tunable qubit is both time- and frequency-dependent, so the resulting time series is multidimensional $P(t,f)$,~\citep{Klimov2018} as in \cref{fig:KlimovFig}.

\subsection{Two-level Systems}

The dominant source of both decoherence and noise in state-of-the-art superconducting quantum circuits is two-level systems (TLS).~\citep{SimLanHitNamPap2004,Muller2019} TLS populations have a large distribution of energies and tunneling rates, and can cause loss in two main ways: through incoherent, off-resonance populations of TLS that are weakly coupled to the device ("incoherent TLS loss"), and through near- or on-resonance coherent individual TLS ("coherent TLS loss"). Interactions between TLS cause fluctuations in induced relaxation and dispersive shifts, leading to parameter fluctuations in qubits ($T_{1}$, $T_{2}^R$, and $f_{01}$) and resonators ($f_{0}$).~\citep{Muller2015,Schlor2019,Bejanin2021}

Incoherent TLS are fluctuators switching incoherently between two eigenstates, and are the dominant source of low-frequency environmental noise affecting both qubits and resonators. Incoherent TLS loss is effectively background loss that is present in the entire 4-10 GHz frequency bandwidth used in superconducting quantum circuits. 

In addition to incoherent TLS, qubits can also couple to coherent TLS located in strong electric field regions, such as Josephson junctions, and resonantly absorb energy from them. This leads to loss at distinct frequencies as well as dispersive shifts which is reflected in $T_1$ and $T_2$ as well as qubit frequency $f_{01}$.~\citep{Schlor2019} For strongly-coupled TLS, anticrossings are induced as shown in Fig.~\ref{fig:SPI}.

\subsection{Resonator Fluctuations Within One Cooldown}

\begin{figure}
\includegraphics[width=85mm]{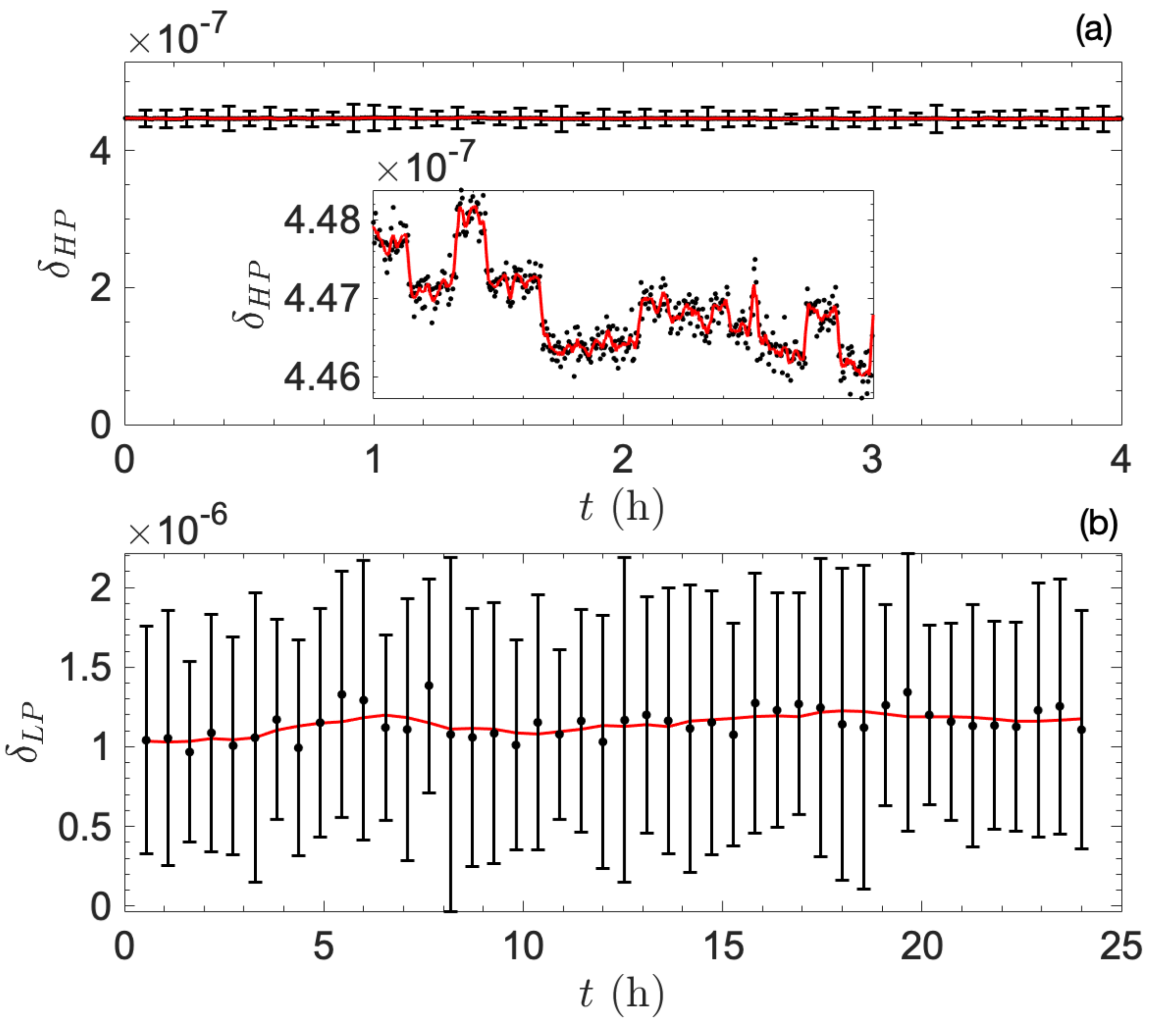}
\caption{\label{fig:restime} Fluctuations of total loss $\delta$ of Nb superconducting microwave resonators at high power over a period of $\sim$ 4 hours (a) and low power over a period of $\sim$ 24 hours (b). Each point in (a) ((b)) is an average of 3 (70) measurements with an IF bandwidth of 1 (0.05) kHz. Red line shows a moving mean over the last 5 (7) data points. Error bars denote the 95$\%$ confidence interval for fit of the resonator transmission data at each timestamp to the Lorentzian-like resonator model.}
\end{figure}

Over a single cooldown, fluctuations in resonator parameters such as resonance frequency $f_0$ and time-dependent phase noise are seen as well as in $Q_i$ (Fig.~\ref{fig:restime}). It is theorized that TLS loss is the single mechanism that contributes to high-frequency fluctuations in performance of planar resonators within a single cooldown.~\citep{Moeed2019}

In addition to TLS-induced noise, sample thermalization over time strongly affects resonator performance. As the sample thermalizes to base temperature, resonator $Q_i$ will decrease, and a shift in resonator frequency $f_0$ will be seen as well. This can be attributed to TLS being saturated at high temperatures but becoming available for device interaction at cooler temperatures.~\citep{Pappas2011,Gao2008b} For accurate results, resonator characterization must occur after full thermalization of the sample to base temperature.

Two techniques are commonly used to perform TLS loss measurements. The first technique measures the loss as a function of power. This can be lengthy (on the scale of many hours) when implementing a power sweep to very low, single-photon powers in order to extract the resonator-induced intrinsic TLS loss. Thus, the power sweep method generally does not capture high-frequency fluctuations of TLS loss. However, TLS loss fluctuations over days or weeks have yet to be studied as well.

The second technique is to probe with a fixed high-power tone and vary temperature. This can decrease the measurement time to minutes per point, with additional time needed for sample thermalization at each temperature.~\citep{Pappas2011} One important caveat is that the temperature sweep method has been demonstrated to give resonator-induced intrinsic TLS loss values that can be 20-30$\%$ higher than the those extracted from the power sweep method. The current understanding of this discrepancy is the temperature sweep method probes a larger frequency range and thus is sensitive to a larger population of TLS.~\citep{Pappas2011} This is an area of study that requires further exploration.

\begin{figure}
\includegraphics[width=80mm]{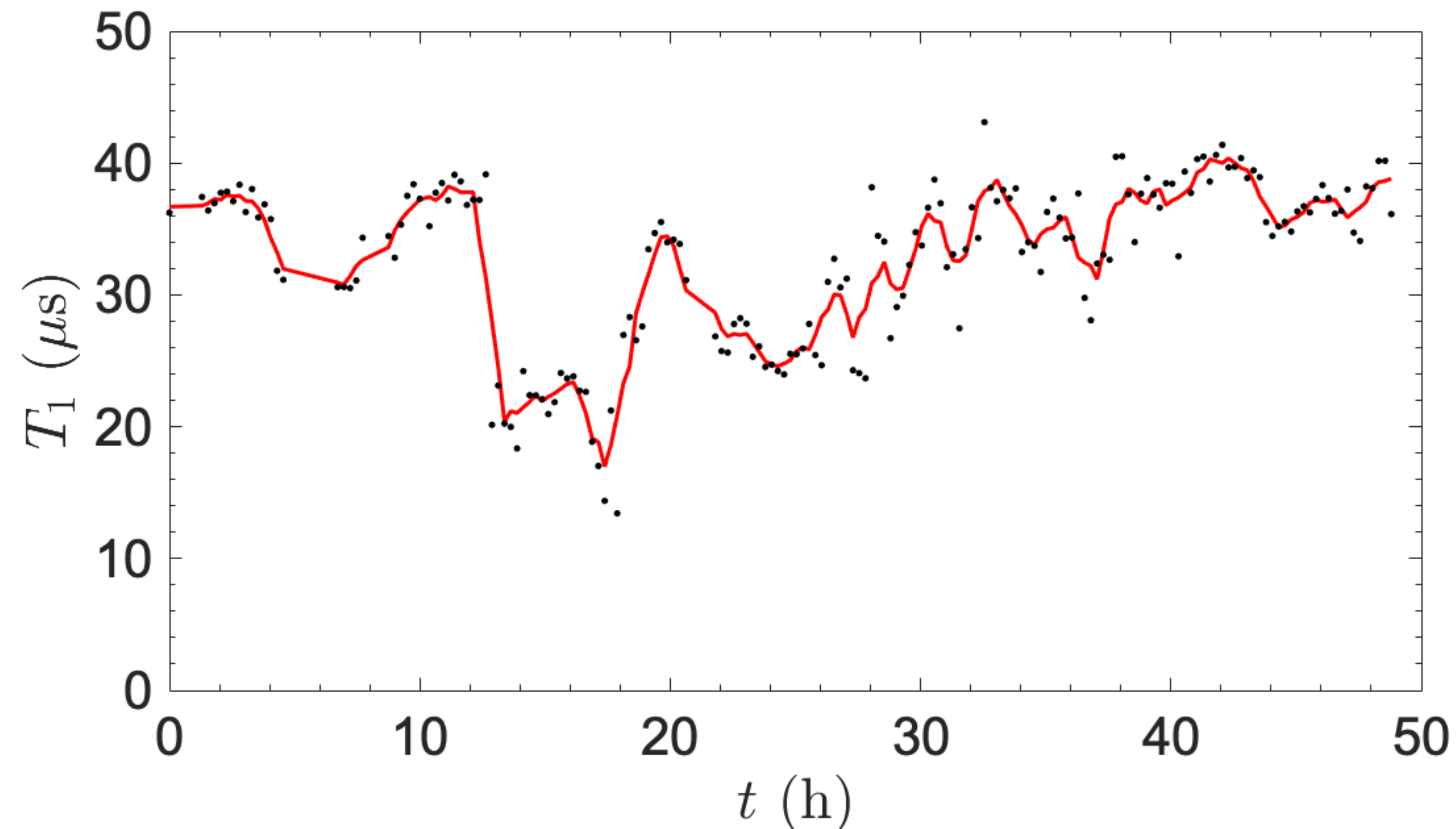}
\caption{\label{fig:qubittime}$T_1$ fluctuations of a grounded concentric transmon qubit over a period of two days. Red line shows a moving mean over the last 5 data points.}
\end{figure}

\subsection{Qubit Fluctuations Within One Cooldown}

\begin{figure}
\includegraphics[width=85mm]{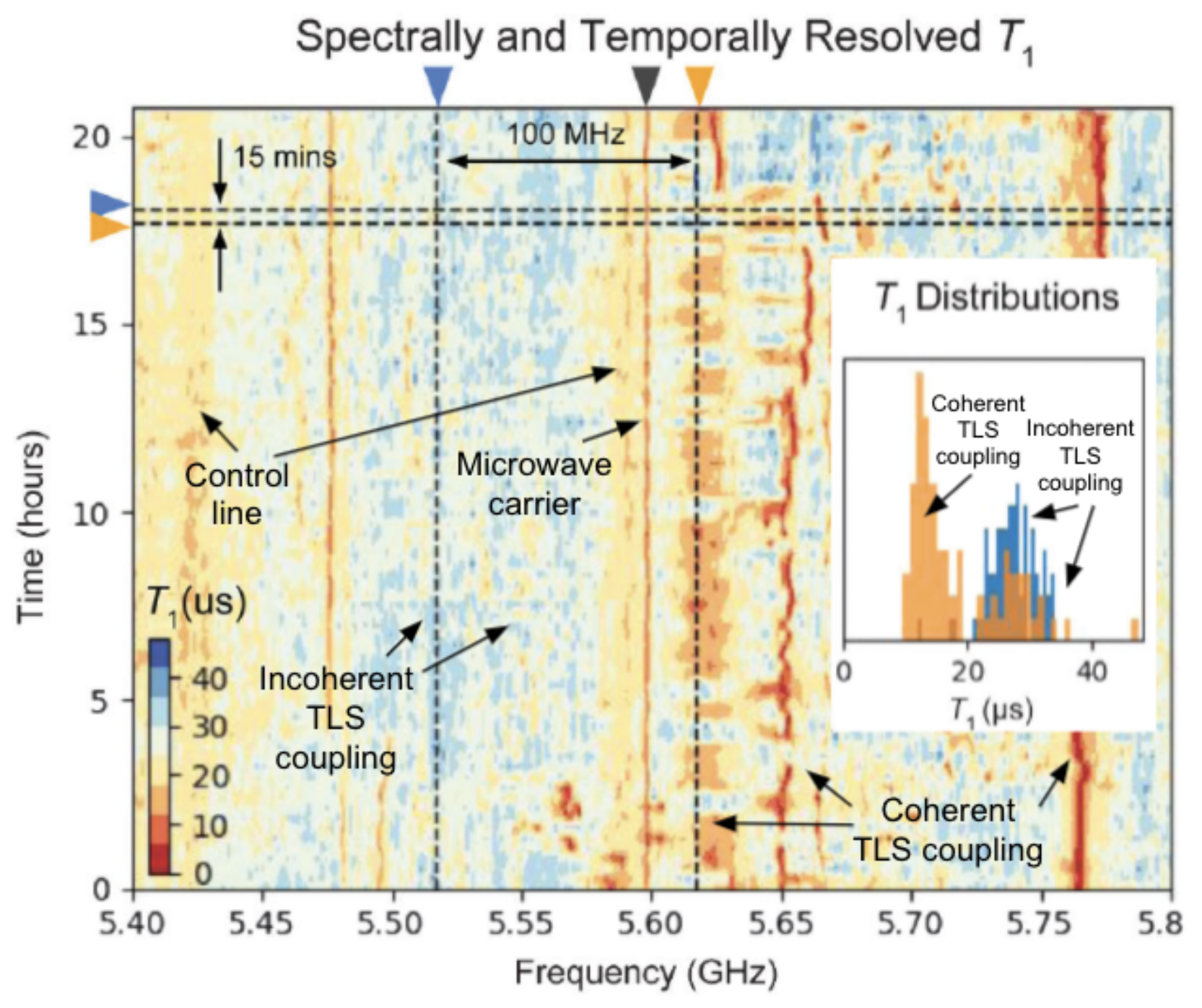}
\caption{\label{fig:KlimovFig}Frequency and time sweeps of a tunable transmon qubit, with labels denoting the source of the low coherence areas. Reproduced with permission from Phys. Rev. Lett. 121, 090502 (2018). Copyright 2018 American Physical Society.}
\end{figure}

Qubit $T_1$ and $T_\phi$, as well as qubit frequency $f_{01}$, vary significantly over time~\citep{Muller2015,Klimov2018,Burnett2019,Schlor2019,Carroll2021} (Fig.~\ref{fig:qubittime}). For $T_1$ in particular, extreme fluctuations can occur on the order of tens of seconds and can be up to an order of magnitude in size.~\citep{Klimov2018} Using evidence from qubit time series and spectral data (Fig.~\ref{fig:KlimovFig}), the largest of these fluctuations can be attributed to individual coherent TLS that move in and out of resonance with the qubit.~\citep{Bejanin2021,Carroll2021,Tersoff2021} A recent study suggests that the few most coherent TLS correspond to more than 10$\%$ of the total loss in standard qubits.~\citep{Tersoff2021} In fact, temporal fluctuations are significant enough that superconducting quantum computing companies periodically take their remotely-accessible devices offline to re-calibrate due to random and systematic drifts in system parameters.~\citep{rigetticomm2,rigetticomm} Indeed, IBM's cloud-accessible quantum processors must be regularly benchmarked and recalibrated in order to manage the effects of these TLS on system performance.~\citep{IBMcomm}

Frequency-tunable qubits can be scanned over both frequency and time, giving a descriptive picture of the spectral environment of the qubit (\cref{fig:KlimovFig}). As well, fixed-frequency qubits can be weakly frequency-tuned by implementing off-resonant microwave tones to drive AC-Stark shifts.~\citep{Carroll2021}

Ref.~\citenum{Carroll2021} indicates that qubit $T_1$ averaged over a nine-month period can be well-described by $T_1$ averaged over a much shorter time period if also implementing tuning over the immediate qubit frequency bandwidth. This suggests that while individual coherent TLS dynamics do strongly affect qubit performance, this effect can be taken into account by probing the near-frequency environment of the qubit. Results in this work indicate that time and frequency bandwidth scan parameters of $\sim$ 1-2 days and $\sim$ 5 MHz can give good correlation to long-term qubit performance.

Frequency scans~\citep{Klimov2018} show sustained lowered $T_1$ at qubit line mode and microwave carrier frequencies. They also indicate frequencies of resonant relaxation that can be attributed to coherent TLS.~\citep{Bejanin2021,Carroll2021,Tersoff2021} When reporting $T_1$ to judge device quality, these low-coherence regions should be avoided so that the incoherent-TLS-loss-limited median $T_1$ can be determined. Coupled with a spectral cleanliness metric, this variable describes the materials-induced decoherence of the device.

Large, correlated errors, such as those due to cosmic rays, can strongly affect superconducting qubit systems.~\cite{Martinis2020} Recent work has shown that cosmic rays can induce correlated, catastrophic qubit decoherence, with events occurring on average every 10 seconds. The effects of these events have been shown to last around 25 ms, with the most impactful of these events leading to $T_1 < 1 \mathrm{\mu}s$.~\cite{McEwen2021} A combination of cosmic ray shielding~\cite{Vepsalainen2020,cardani2020reducing} and on-chip mitigation could solve this issue, but further investigation is required.

Little information is currently published on temporal cross-correlation or causal relations between performance metrics. This is an important area for future research.

In opposition to resonator performance, qubit $T_1$ increases as the sample decreases in temperature.~\citep{Lisenfeld2007,Lisenfeld2010} Thus, the confirmation of sample thermalization is necessary to ensure an optimal measurement environment.

The temperature variation of ambient-temperature electronics and microwave components, over time and even by time of day has been anecdotally shown to affect qubit performance.~\citep{rigetticomm2,rigetticomm} The measurement of devices over long time periods can help avoid systematic, uncontrolled biasing of the measurement result.

\section{Cooldown-to-Cooldown Variation}\label{sec:cooldown}
The on-chip TLS configuration, including the frequencies and number of coherent TLS, has been observed to change with each cooldown,~\citep{Burnett2019} implying a "freeze out" effect where the population of coherent TLS in a device vary significantly between cooldowns but not within a single cooldown. What remains unclear is whether median time series data and performance metrics within a cooldown are consistent with subsequent cooldown data. Here, we define a subsequent cooldown to be cooling a sample to base temperature from ambient temperature without opening the dilution refrigerator. 

The primary questions are: (1) Are the changes in the distributions of the performance metrics statistically significant between cooldowns? and (2) If so what are the magnitudes of these changes? Currently, the type and magnitude of these cooldown-to-cooldown changes of performance metrics are largely unstudied.

Across cooldowns, resonator $Q_i$ at high power has been shown to vary by more than 50$\%$, whereas TLS loss remains within the fit uncertainty.~\citep{Calusine2018} This is consistent with large cooldown-to-cooldown variations in experimental environment (which is a strong factor in high power $Q_i$) and negligible cooldown-to-cooldown variation in the incoherent TLS population.

\section{Effects of Microwave Packaging}\label{sec:mwpackage}
The microwave packaging that houses a superconducting circuit is generally transferable between systems and can thus be seen as independent of other setup-to-setup variations. This distinction becomes important when implementing interlaboratory comparisons of devices. Wirebonds, lossy materials, device thermalization, and box modes are all factors that can affect the perceived performance of superconducting quantum circuits.~\citep{Wenner2011,Rosenberg2019,Lienhard2019,Mergenthaler2020,Huang2021}

Wirebond inductance causes loss-inducing crosstalk that increases with length, whereas stray transmission can be suppressed by increasing the density of grounding wirebonds.~\citep{Wenner2011} A high density of short grounding wirebonds is commonly implemented to balance these conflicting effects, but more research is required to determine decoherence effects of wirebond repeatability in terms of variation in length and placement. Alternatives such as pogo pins~\citep{Bronn2018,Bejanin2016}, air bridges,~\citep{Abuwasib2013,dunsworth2018_airbridge,chen2014_airbridge} and bump bonds~\citep{Rosenberg2017,Foxen2018} can also be implemented in some instances and may allow greater experimental reproducibility.

When present, geometry-dependent box modes can couple to on-chip devices and cause circuit decoherence with nebulous behavior. Ideally, box modes should be far detuned from the spectrum of interest~\citep{Wenner2011,Mcconkey2017,Lienhard2019} in order to facilitate measurement reproducibility between sample boxes. A recent study suggests that package modes due to the Purcell effect can limit the qubit lifetime to $T_1 \lesssim 384 \mathrm{\mu}s$.~\cite{Huang2021} Box modes can also facilitate crosstalk between on-chip devices, leading to coherent errors and adding another layer of complexity to performance comparisons. 3D integration such as bump bonds and vias can allow greater control over the microwave environment of on-chip devices.~\citep{Rosenberg2019}

Sample boxes are often made from high purity Al or gold-plated Cu.~\citep{Lienhard2019} Cu boxes have strong thermalization properties but also generate loss due to normal metal conduction,
although in some packaging designs this loss is orders of magnitude lower than other losses affecting qubits and resonators.~\citep{Lienhard2019} On the other hand, Al boxes are superconducting and thus do not introduce normal conducting loss, but are poor thermal conductors. 

In addition, lossy materials such as dielectrics and normal metals are sometimes included in the vicinity of the chip, such as on circuit boards or as a paste to thermalize and adhere the sample to the box. These components are lossy and, in the case of thermal paste, are not necessarily reproducible to implement. The replacement of these components by constant or lower-loss substitutes, such as the implementation of superconducting circuit boards~\citep{Pappas2018} and clamps rather than paste, is likely to improve measurement repeatability with changing microwave packaging.

\section{Setup-to-Setup Variation}\label{sec:setup}
Performance metric measurements of superconducting devices are heavily influenced by attenuation, filtering, amplifiers, wiring, and other components between the ambient-temperature electronics and the superconducting circuits.~\citep{ranzani2013two, Wang2021} Connections to ambient-temperature control electronics influence both qubit relaxation and pure dephasing.~\citep{Yeh2017,Yan2016,Ithier2005,Krinner2019} Noise properties can vary significantly even between sources from the same manufacturer. In addition, the magnitude of the passive infrared radiation load can vary between dilution refrigerator set-ups. Together, these factors add to the complexity of reproducing performance-enhancing results between systems.

In qubit frequency sweeps such as those in Ref.~\citenum{Klimov2018} (Fig.~\ref{fig:KlimovFig}), distinct frequencies of consistent low coherence can be identified. These frequencies can be attributed to factors in the experimental chain such as control line modes and microwave carrier signal bleedthrough, and will therefore differ between setups. The identification and subsequent detuning of the qubit from these frequencies is critical to allow comparisons of materials-limited qubit coherence as well as to optimize qubit performance.

For resonator characterization, cryogenic calibration such as in Refs.~ \citenum{ranzani2013two} and \citenum{Wang2021} can account for the vast majority of the differences between experimental setups, save for those past the calibration plane which normally includes microwave packing and at least one connector. Additionally, hanger-mode resonator fitting is self-calibrating~\citep{Khalil2012, Probst2015, GaoThesis} and thus changes in the microwave chain have little effect on the extracted $Q_i$ of hanger-mode resonators but a strong impact on reflection- and transmission-mode resonators.

Sample thermalization, a critical factor in measurement reproducibility, can be affected by experimental setup as well as sample packaging. The connection between the microwave packaging and base plate as well as thermalization between temperature stages determines in part the thermalization rate of the sample. Inefficient thermalization can lead to elevated sample temperatures and thus reduce qubit performance and elevate resonator performance. Thermalization of superconducting devices in dilution refrigerators has yet to be studied in depth, but a good thermal link and method for accurate sample temperature measurement would sufficiently address this concern. Possible thermometry methods include implementing the qubit itself as a thermometer.

\section{Cross-Wafer Variation and Variation Between Fabrication Runs}\label{sec:faberror}
Nanofabrication processes for superconducting quantum devices are still largely performed in academic or small-scale cleanrooms that do not have the same process control or device reproducibility as large-scale semiconductor foundries. Industrial efforts in superconducting quantum device manufacturing have likely produced in-depth but unpublished coherence reproducibility studies. Thus, fabrication-induced coherence inhomogeneity across the wafer and between fabrication runs remain as significant unknowns. Variation in factors such as the superconducting transition temperatures, film thicknesses, etch rates, mechanical stress, and oxidation could have strong but currently unstudied effects on device performance.

As well, some of these small fabrication differences are believed to have a significant impact on TLS populations and therefore device performance. Indeed, resonators on the same chip have been shown to have $Q_i$ that vary by up to an order of magnitude.~\citep{Ohya2014}

As outlined in Ref.~\citenum{Woods2019}, on the order of 120 resonators are needed in order to extract individual losses from metal-substrate, substrate-air and metal-air interfaces as well as the substrate using the surface loss extraction technique.~\citep{Calusine2018} We posit that this can be thought of as an extreme upper bound for the number of resonators needed to capture performance fluctuations due to fabrication for a single device design and fabrication process.

Cross-wafer and fabrication-run variation in performance are especially difficult to characterize as these studies require taking into account errors in all other aspects of the experiment. This is the most nebulous of all the variation types seen in superconducting quantum circuit experiments. For this reason, ambient-temperature characterization methods that correlate to millikelvin microwave device performance are seen as a holy grail in superconducting quantum computing.

\begin{table*}
\caption{\label{tab:res}Summary of error types that affect $Q_i$ and $F \delta_{TLS}^0$ measurements in planar superconducting microwave resonators, including error sources, typical error magnitudes, and possible solutions for minimizing or accounting for error. TBD: further investigation is required. $\sigma_{A}$: standard deviation of $A$. $\Delta A$: variation in A.}
\begin{ruledtabular}
\begin{tabular}{cccc}
\textbf{Type} & \textbf{Source} & \textbf{Magnitude} & \textbf{Solution} \\
\hline
Measurement $\&$ analysis & $Q_i$ fit error & TBD & Design for low error regime \\
 & $Q_i$ fit uncertainty & $\sigma_{Qi} \pm \sim 2\%$ (20$\%$) for low (high) noise~\footnote{Typical values based on past measurements of Al hanger mode CPW resonators at the Boulder Cryogenic Quantum Testbed, University of Colorado Boulder.} & Report uncertainty \\
 & TLS fit uncertainty & $\sigma_{TLS} \sim \pm 5\%$ for power sweep~\citep{McRae2021} & Report uncertainty, reduce low power noise \\
\hline
Within one cooldown & Incoherent TLS & $\Delta Q_i \lesssim 30 \%$ over several hours~\citep{Megrant2012,neill2013fluctuations} & Time average data \\
&  & $\sigma_{Qi} \sim$ 12$\%$ over three hours ~\citep{Moeed2019} \\
& Thermalization over time & $\Delta Q_i \lesssim$ orders of magnitude~\citep{Barends2011,McRae2020} & Ensure sample thermalization \\
&  &  & prior to measurement \\
\hline
Cooldown-to-cooldown & Microwave environment & $\Delta Q_{i,HP} > 50 \%$~\citep{Calusine2018} & Report $\delta_{TLS}^0$ rather than low power $Q_i$ \\
 & & $\Delta \delta_{TLS}^0 \sim 0$~\citep{Calusine2018} \\
\hline
Microwave packaging & Wirebonds & Varies & High density of low wirebonds,~\citep{Wenner2011} \\
 &  &  & alternative solutions~\citep{chen2014_airbridge,Bejanin2016,Bronn2018,,Abuwasib2013,dunsworth2018_airbridge,Rosenberg2017,Foxen2018} \\
 & Sample box material & Varies & TBD \\
 & Box modes & Varies & Design to suppress box modes \\
 & Lossy materials & Varies & Use lower-loss substitutes \\
\hline
Setup-to-setup & Microwave chain & Can lead to unphysical values in $Q_i$~\citep{Wang2021} & Calibration or hanger-mode~\citep{ranzani2013two,Wang2021} \\
 & Thermalization & Varies, can elevate $Q_i$ & Good thermal link,  \\
  &  &  & $T$ measurement at sample \\
\hline
Cross-wafer & Resist, deposition, & Unknown & Measure a large number of \\
 &  etch inhomogeneity & & nominally identical devices \\
 \hline
Between fabrication runs & Changes in chamber & Unknown & Measure a large number of \\
 &  particulate, etch rates, etc. & & nominally identical devices \\
\end{tabular}
\end{ruledtabular}
\end{table*}

\begin{table*}
\caption{\label{tab:qubit}Summary of error types in $T_1$ and $T_\phi$ measurements of superconducting transmon qubits, including error sources, magnitudes, and possible solutions for minimizing or accounting for error. TBD: further investigation is required. $\sigma_{A}$: standard deviation of $A$. $\Delta A$: variation in A.}
\begin{ruledtabular}
\begin{tabular}{cccc}
\textbf{Type} & \textbf{Source} & \textbf{Magnitude} & \textbf{Solution} \\
\hline
Measurement and analysis & Error in TLS detection & TBD & Automate detection process \\
\hline
Within one cooldown & Incoherent TLS & $\Delta T_1 \sim 20\%$~\citep{Klimov2018} & Time average data \\
& Coherent TLS & $\Delta T_1 \lesssim$ order of magnitude~\citep{Klimov2018} & Measure over time and frequency~\cite{Carroll2021} \\
& Cosmic rays & $\Delta T_1 \lesssim$ order of magnitude~\citep{McEwen2021,Martinis2020} & Shielding,~\cite{Vepsalainen2020,cardani2020reducing} on-chip mitigation \\
& Thermalization over time & Varies~\citep{Lisenfeld2007,Lisenfeld2010} & Ensure sample thermalization \\
&  &  & prior to measurement \\
& Temperature of RT electronics & TBD & Periodic recalibration, \\
&  &  & temperature control \\
\hline
Cooldown-to-cooldown & Freeze-out of TLS & $\Delta T_1 \lesssim$ order of magnitude~\citep{Klimov2018} & Detune from TLS \\
\hline
Microwave packaging & Wirebonds & Varies & High density of low wirebonds,~\citep{Wenner2011} \\
 &  &  & alternative solutions~\citep{Bronn2018,Bejanin2016,Abuwasib2013,dunsworth2018_airbridge,chen2014_airbridge,Rosenberg2017,Foxen2018} \\
 & Sample box material & Varies & TBD \\
 & Box modes & Varies & Design to suppress box modes \\
 & Lossy materials & Varies & Use lower-loss substitutes \\
\hline
Setup-to-setup & Microwave chain & Very large~\citep{Krinner2019,Klimov2018} & Detune from regions \\
 & Thermalization & Varies, can reduce $T_1$ & Good thermal link, \\
   &  &  & $T$ measurement at sample \\
\hline
Cross-wafer & Resist, deposition, & Unknown & TBD \\
 &  etch inhomogeneity \\
 \hline
Between fabrication runs & Changes in chamber & Unknown & TBD \\
 &  particulate, etch rates, etc. \\
\end{tabular}
\end{ruledtabular}
\end{table*}

\section{Reproducibility Guidelines}\label{sec:guidlines}
In this section, we provide a non-exhaustive summary of best practices as well as current estimates for orders of magnitude of variation of performance metrics for qubits and resonators (Tables~\ref{tab:res} and \ref{tab:qubit}). These guidelines could allow reproducibility and one-to-one comparisons in interlaboratory measurements, which could enable true breakthroughs in superconducting quantum device performance.

\subsection{Data guidelines:}

\begin{itemize}
    \item In order to facilitate future data reanalysis based on new understandings, as well as one-to-one comparison, make data publicly available in common file formats such as \verb!csv! or \verb!JSON! and include sufficient metadata such as powers, temperatures, and number of averages as applicable.
    \item Make data analysis routines version-controlled and publicly available, and state which version of analysis routines were used.
    \item Provide order-of-magnitude error estimates, even if all error sources were not explicitly probed. Include fit uncertainties.
\end{itemize}

\subsection{Experimental guidelines:}

\begin{itemize}
    \item Interleave measurements of different performance metrics when possible to enable cross-correlation analysis.
    \item Record laboratory and control electronics temperature to enable cross-correlation analysis with the performance metrics.
    \item Design packaging to suppress box modes and substitute lossy materials near the sample to allow measurement repeatability between microwave packages.
    \item For resonator characterization, hanger-mode geometry should be implemented, or cryogenic calibration measurements should be performed to account for the effects of the microwave chain between the vector network analyzer and the sample.
    \item Characterize qubit performance over a time span of at least 1-2 days and a small frequency bandwidth around the idle frequency of the qubit in order to capture coherent TLS effects.
    \item For a more complete qubit characterization, include a spectral cleanliness metric along with median $T_1$ and $T_\phi$ at an incoherent-TLS-loss-limited frequency or at the idle frequency of the 
    qubit.
    \item Measure more than one nominally identical device when possible in order to capture some element of fabrication-induced performance variation.
    \item Perform multiple cooldowns when possible in order to capture changes in coherent TLS population when cycling to room temperature.
\end{itemize}

\subsection{Reporting guidelines:}

\begin{itemize}
    \item Include description of entire experimental set-up, including microwave wiring, sample mount, and microwave packaging.
    \item Include description of device design, including material filling factors when possible.
    \item Report median and an inter-percentile as summary statistics of the times series data. These are robust statistics of central tendency and spread, unlike the mean and standard deviation. 
    \item Include accurate sample temperature data when possible.
\end{itemize}

\section{Expanding the Frontier of Reproducible Coherence Characterization}

In previous sections, we outline known sources of error and variation in superconducting quantum device measurement, and we recommend best practices for reproducible coherence characterization of these devices. In this section, we highlight areas for knowledge expansion and propose experiments to heighten our understanding in these areas.

One important goal is to increase understanding of loss sources, fluctuations of loss, and magnitudes of loss in superconducting devices. Measuring both longer- and shorter-scale time series of resonator $Q_i$ and correlating them with qubit $T_1$ could provide this additional information. Particularly, longer measurements would allow the investigation of fluctuations in performance due to changes in ambient temperature and coherent TLS frequency drift. In a similar vein, the development of a rapid TLS measurement protocol would allow investigations of TLS loss fluctuations over time. 

The comparison of fluctuations of low-power and high-power $Q_i$ could allow us to more easily distinguish between types of loss. 

It is critical understand the freeze-out effect in both incoherent and coherent TLS populations. Such effects can be studied by cooldown-to-cooldown variation of both qubits and resonators, where devices are cycled from ambient to millikelvin temperatures many times. Of course, the reported data would conform to the guidelines laid out in \cref{sec:guidlines}. 

Accurate temperature measurements of the sample itself rather than the mixing chamber stage would allow investigations of thermalization, both of the sample to the microwave packaging and the package to the base plate.

A comprehensive study of microwave packing, including geometry, bulk material, circuit board, and thermalization method, must be performed in order to understand the full extent of these effects. 

The investigation that relies on reproducible coherence characterization in all other aspects is the investigation of fabrication method effects. This set of experiments will allow us to optimize superconducting quantum device performance and could possibly enable large-scale superconducting quantum computing in the future. An attractive first step would be the implementation of rapid resonator characterization to allow the collection of larger datasets.

Finally, multiqubit benchmarking, while not included in this work, is essential to the investigation of correlated and spatially-dependent performance fluctuation sources such quasiparticles due to cosmic rays,~\cite{Martinis2020,McEwen2021} and further investigations into spatially-dependent performance correlations is warranted.

\section{Conclusion}
In conclusion, we have explored error contributions to superconducting device performance metrics from various experimental areas, including measurement and analysis error, fluctuations over time, cooldown-to-cooldown variations, effects of microwave packaging, setup-to-setup variations, and fabrication-induced variations. Experimental areas requiring further study were elaborated upon, and experimental guidelines for reducing and accounting for these errors were noted.

This article {\em is not} intended to cast aspersions on the reported results in the literature. Steady progress in superconducting quantum device performance has been observed for decades in academic and industrial labs all over the world, in contrast to other qubit modalities.~\citep{frolov2021quantum} Large improvements in device performance, especially those due to design modification, have been largely reproducible from one laboratory to another. That is not to say that all improvements are comparable or even reproducible. As the field matures and reported improvements become more modest, it is difficult to be certain the results are reproducible. 

Looking to the future of the field, there are a number of paths towards reproducible coherence characterization. The simplest path is to report experimental details regarding known variation sources, which would enable interlaboratory performance improvements. On the more ambitious end is the development of a standard, optimized, open-source dilution refrigerator for coherence characterization (including standardized wiring, control electronics, mounts, data acquisition and analysis software, and commissioning and verification procedures). Such standard experimental set-up could then be straightforwardly reproduced.

\begin{acknowledgments}
The authors would like to thank Clemens Müller, Stefano Poletto, Alexa Staley, Michael Vissers, and Ruichen Zhao for helpful discussions and feedback on the manuscript.

This material is based upon work supported by the U.S. Department of Energy, Office of Science, National Quantum Information Science Research Centers, Superconducting Quantum Materials and Systems Center (SQMS) under the contract No. DE-AC02-07CH11359.
\end{acknowledgments}

\section*{Data Availability}
The data that support the findings of this study are openly available in Boulder-Cryogenic-Quantum-Testbed/data at \url{https://doi.org/10.5281/zenodo.4902377}.

\section*{References}
\bibliography{APLperspective}

\end{document}